\documentclass[twocolumn]{emulateapj}

\received{2017 September 02}
\revised{2017 October 30}
\accepted{2017 November 8}

\shorttitle{Star formation near R\,136}
\shortauthors{Kalari et al.}

\begin{document}

\title{Pillars of creation amongst destruction: \\ Star formation in molecular clouds near R\,136 in 30 Doradus  }

\email{venukalari@gmail.com}

\author{Venu M. Kalari}
\affil{Departamento de Astronomia, Universidad de Chile, Casilla 36-D, Santiago, Chile}
\author{M{\'o}nica Rubio}
\affil{Departamento de Astronomia, Universidad de Chile, Casilla 36-D, Santiago, Chile}
\author{Bruce G. Elmegreen}
\affil{IBM Research Division, T.J. Watson Research Center, 1101 Kitchawan Road, Yorktown Heights, NY 10598 USA}
\author{Viviana V. Guzm{\'a}n}
\affiliation{Joint ALMA Observatory (JAO), Alonso de Cordova 3107 Vitacura, Santiago, Chile}
\author{Cinthya N. Herrera}
\affil{Institut de Radioastronomie Millimétrique, 300 rue de la Piscine, Domaine Universitaire, 38406, Saint-Martin-d'Hères, France}
\author{Hans Zinnecker}
\affil{Deutsches SOFIA Institut (DSI) University of Stuttgart, Pfaffenwaldring 29, D-70569, Germany}
\affil{Universidad Aut{\'o}noma de Chile, Av. Pedro de Valdivia 425, Santiago, Chile}


\begin{abstract}
New sensitive CO\,(2-1) observations of the 30 Doradus region in the Large Magellanic Cloud are presented. We identify a chain of three newly discovered molecular clouds we name KN1, KN2 and KN3 lying within 2--14\,pc in projection from the young massive cluster R\,136 in 30 Doradus. Excited H$_2$\,2.12$\mu$m emission is spatially coincident with the molecular clouds, but ionized Br$\gamma$ emission is not. We interpret these observations as the tails of pillar-like structures whose ionized heads are pointing towards R\,136. Based on infrared photometry, we identify a new generation of stars forming within this structure. 


\end{abstract}

\section{Introduction} \label{sec:intro}

30 Doradus is a giant H{\scriptsize II} region in the Large Magellanic Cloud (LMC). The LMC is a local group dwarf galaxy that lies at a distance of 50\,kpc (Pietrzy{\'n}ski et al. 2013), and has a mean stellar metallicity ($Z$) half of the Sun (Rolleston et al. 2002). 30 Doradus hosts the young massive cluster (YMC) R\,136. R\,136 is a $\sim$1.5-3\,Myr YMC that encloses a total cluster mass in excess of 10$^5$\,$M_{\odot}$ within 10\,pc (Selman \& Melnick 2013). The cluster contains roughly 200 massive stars ($>$8\,$M_{\odot}$) within a central region less than 6\,pc, whose radiation and mechanical feedback profoundly impact the surrounding medium (Schneider et al. 2017, subm.). R\,136 is the most massive YMC in our local neighbourhood that can be adequately resolved spatially (at 50\,kpc, the nominal distance to R\,136, 1$\arcsec$$\approx$0.25\,pc) enabling us observe individual objects at the star and molecular clump scale. This makes R\,136 an ideal laboratory to examine how feedback from massive stars affects further star formation (e.g. Dale et al. 2012). 

\begin{figure*}
	\plotone{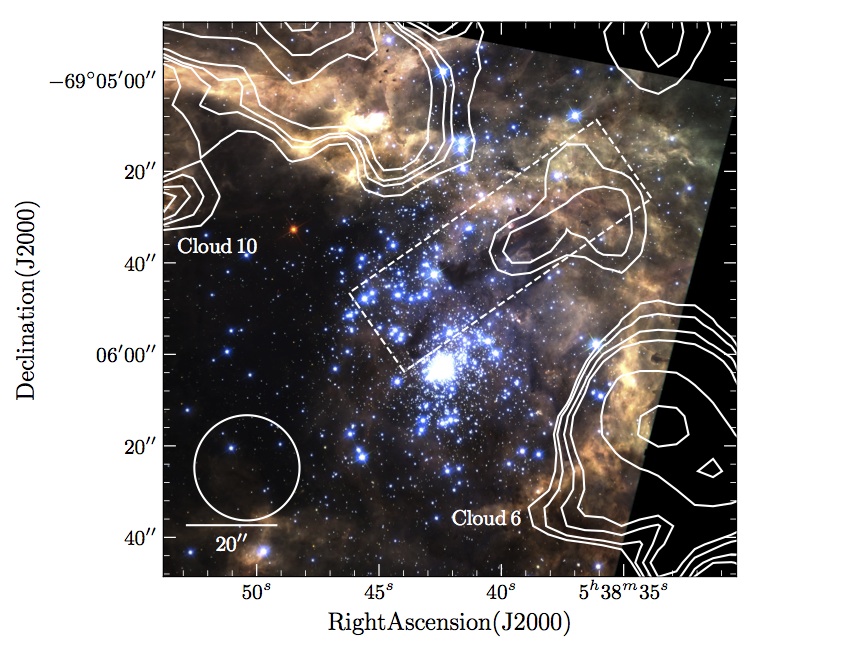}
\caption{$B/V/I$H$\alpha$ (=b/g/r) three colour HST image of 30 Doradus, spanning 2$'\times2'$, centred near R\,136. North is up and east is to the left, with the scalebar on the lower left indicating 20$''$. At the distance to the LMC, 20$''\approx$5\,pc. R\,136 is visible as the bright star cluster at $\alpha$=05$^h$42$^m$ and $\delta$=-69$\degr$06$'$. The white dashed rectangle marks the boundary of the stapler nebula discussed in this work.  CO\,(2-1) contours are shown in white, with the beam size given by the lower left circle. The emission is integrated over the velocity range of 235--270\,km\,s$^{-1}$. Contour levels are from 0.8--2.4\,K\,km\,s$^{-1}$ in steps of 0.4\,K\,km\,s$^{-1}$ and from 3.2--10\,K\,km\,s$^{-1}$ in steps of 0.8\,K\,km\,s$^{-1}$.}
    \label{fig:CO}
\end{figure*}

The mechanical and radiation output from R\,136 has created a central cavity by sweeping the surrounding molecular clouds (labelled as clouds 6 and 10 in Fig.\,1) that extend up to 100\,pc along the northeast-southwest axis (Pellegrini et al. 2010). We adopt the cloud nomenclature of Johansson et al. (1998). Brightly illuminated arcs delineate the interfaces between the cold gas and the ionizing radiation, where subsequent generations of stars are thought to have been triggered (Walborn et al. 2002). Studies at optical (De Marchi et al. 2011; Kalari et al. 2014), near-infrared (nIR; Rubio et al. 1998; Brandner et al. 2001), mid-infrared (mIR; Whitney et al. 2008; Gruendl \& Chu 2009; Walborn et al. 2013), far-infrared (fIR; Seale et al. 2014) and sub-millimeter (Johansson et al. 1998; Indebetouw et al. 2013) wavelengths have identified evidence for active star formation throughout the 30 Doradus nebula, consistent with the idea of multiple star formation episodes. 


We focus on the stapler nebula that lies 2-14\,pc away from R\,136 (see Fig.\,1) in projection. The stapler nebula is the H{\scriptsize II} region including and surrounding the stapler shaped dark cloud that is seen in silhouette in the optical near R\,136. The nebula spans an area of 1.1$'\times$0.35$'$ centred on $\alpha$\,=\,05$^h$38$^m$40$^s$, $\delta\,=\,-$69$^{\rm \circ}$05$'$36$''$ and is elongated with a position angle of 35$^{\rm \circ}$. A candidate young stellar object (YSO) has been reported at the edge of the elongated dark cloud by Walborn et al. (2013; marked as S5 in that paper). The YSO is close to, but not coincident with a region of high density ($n>$10$^{6}$\,cm$^{-3}$) reported by Rubio et al. (2009) using CS line observations. Known infrared excess objects, some of which are thought to be disc/envelope bearing young stellar objects (YSOs) are dotted towards the edge of the dark cloud according to Rubio et al. (1998; their Figure 3). The literature evidence for dense molecular gas and YSOs in the stapler nebula lying near R\,136 indicates that star formation may be ongoing, which deserves further study. In this paper we discuss the properties of molecular clouds in the stapler nebula, and examine whether new stars are being formed in these clouds.  


This paper is organised as follows. In Section 2 we describe the data used in this study. The results from the analysis of CO\,(2-1) line observations are presented in Section 3. We discuss the results obtained from nIR emission line images of the stapler nebula in Section 4. Based on archival infrared photometry, we identify YSOs within the stapler nebula in Section 5. The picture obtained from our results is described in Section 6. In Section 7, a brief summary of our paper is presented along with future work arising from our results.

\begin{figure*}
\includegraphics[width=0.495\textwidth]{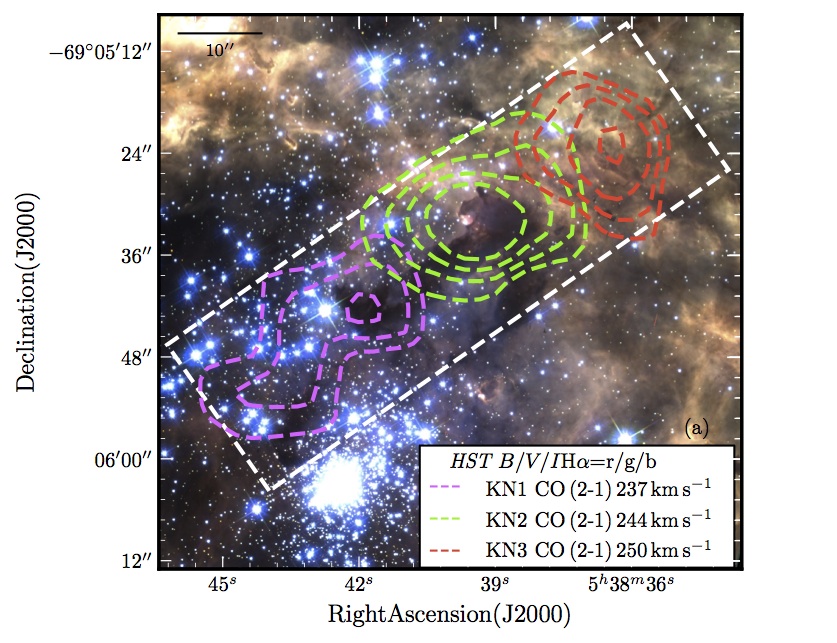}
\includegraphics[width=0.495\textwidth]{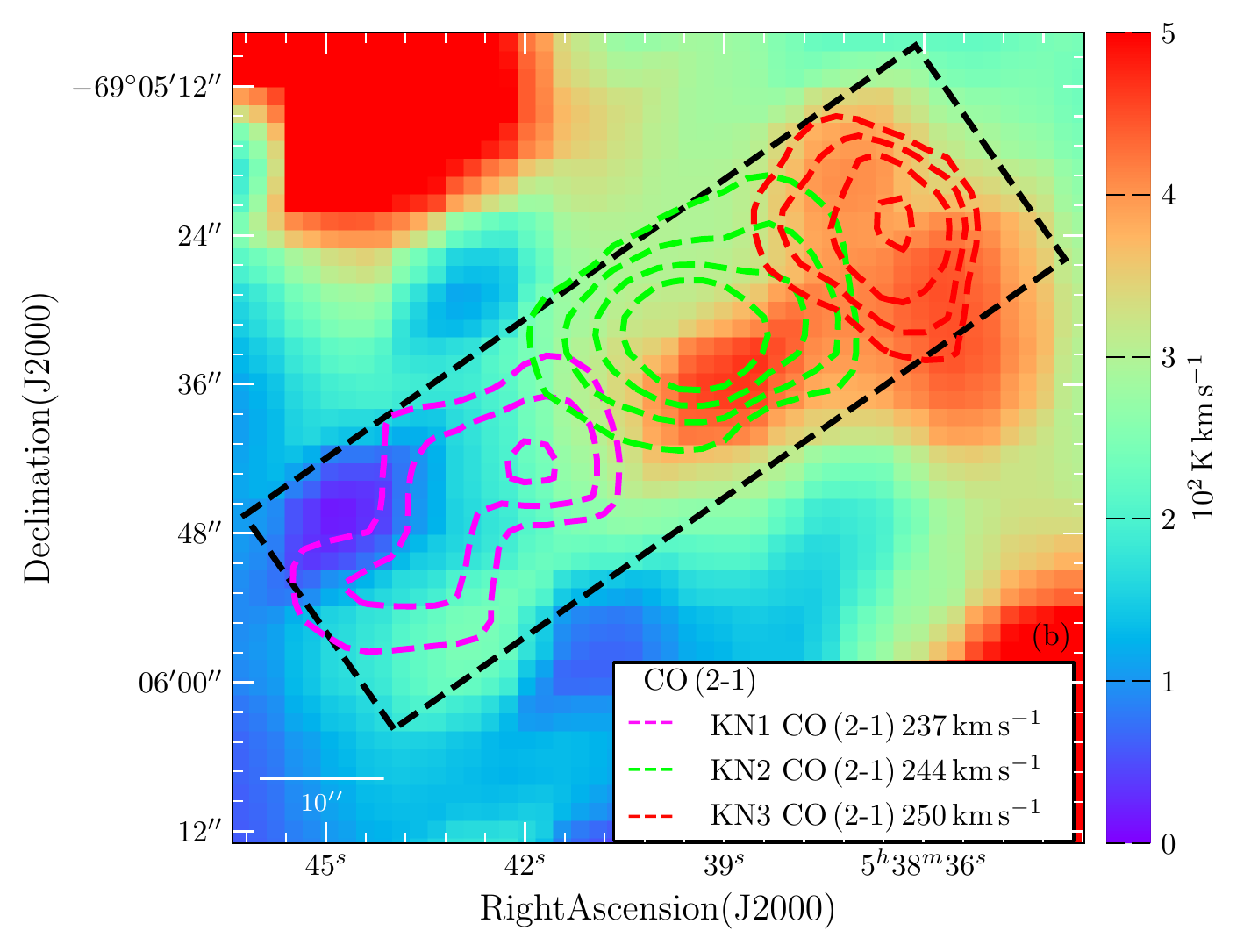}
\includegraphics[width=0.495\textwidth]{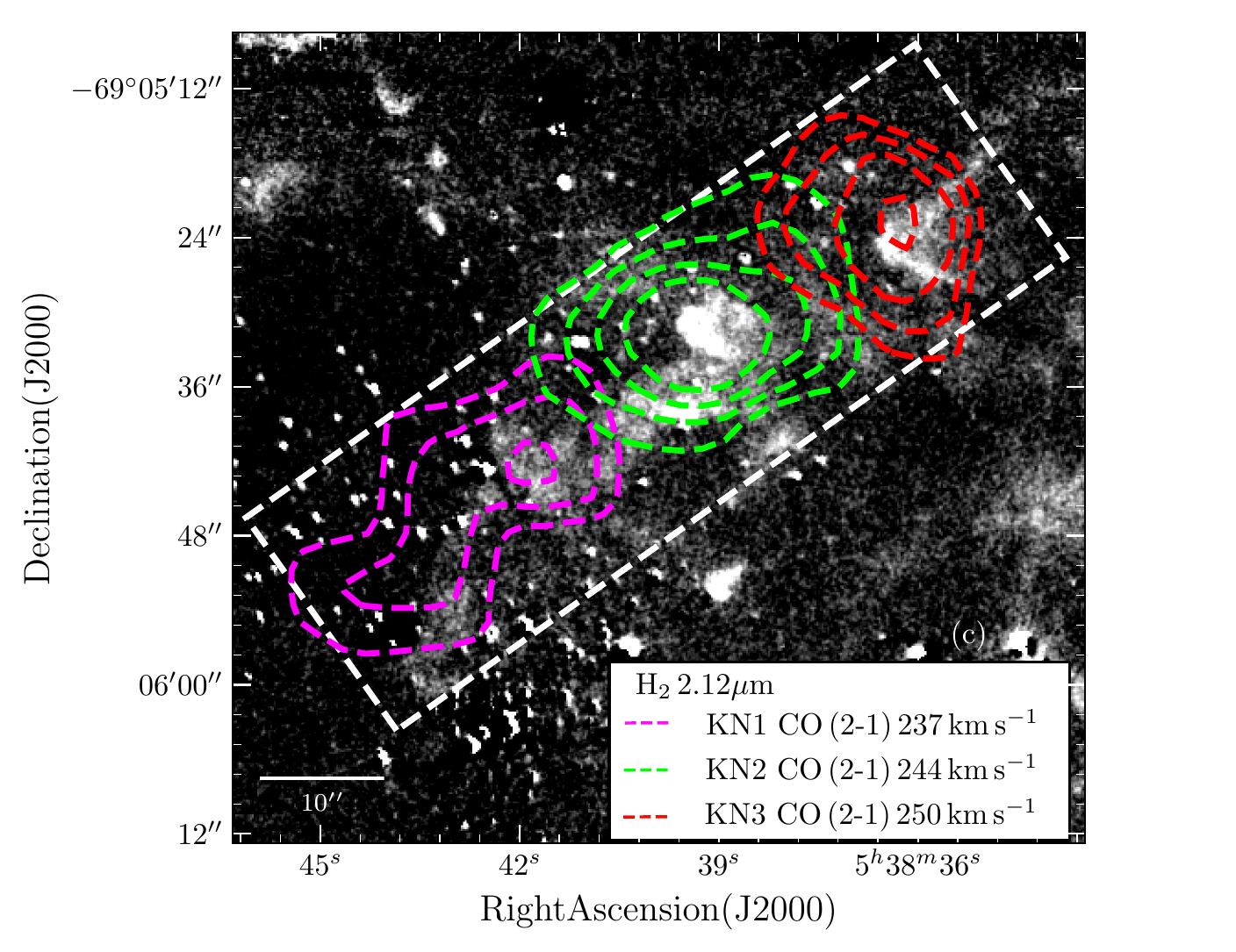}
\includegraphics[width=0.495\textwidth]{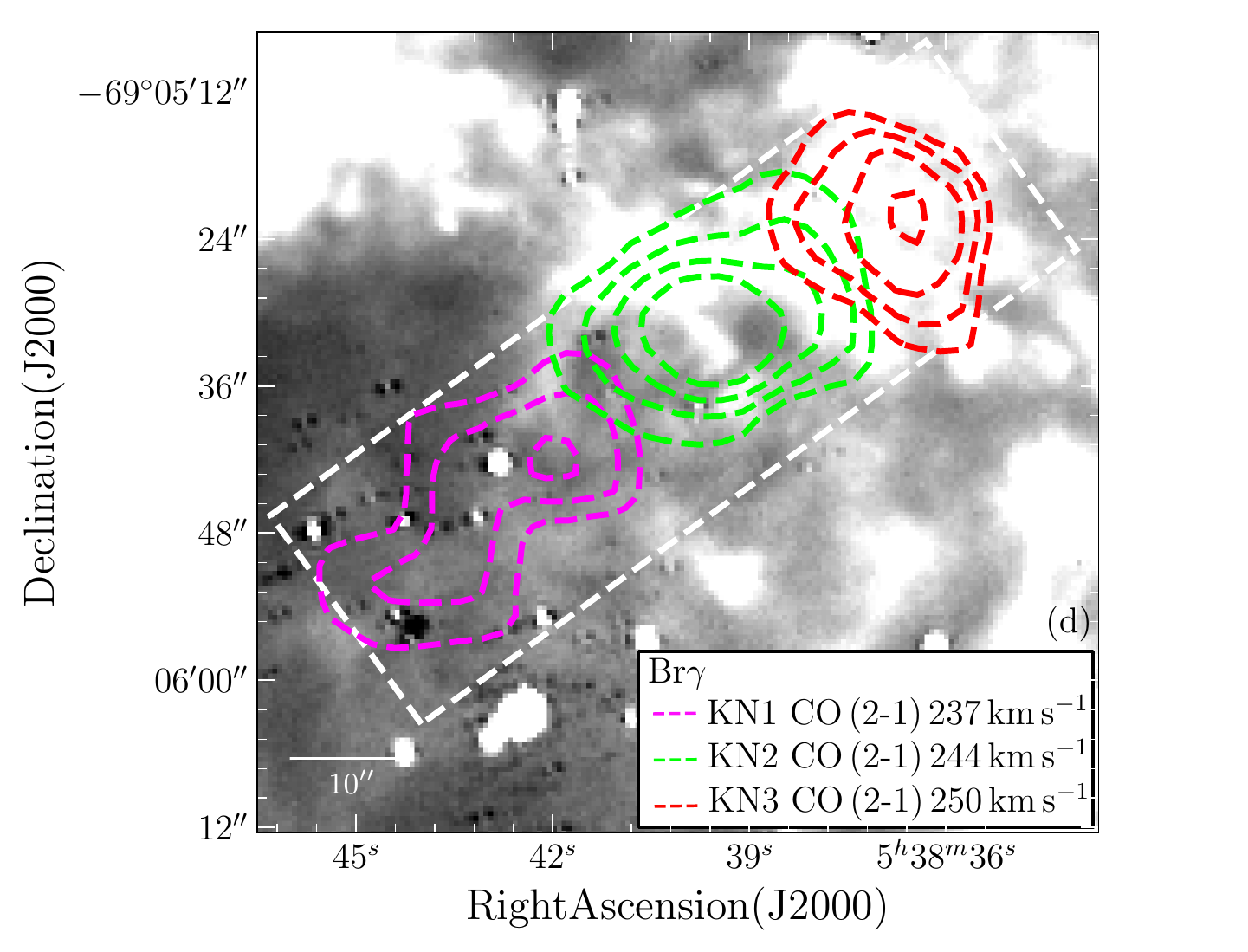}
\caption{(a)The stapler nebula blown up from the HST mosaic in Fig.\,1. The stapler region is outlined with the dashed white box. The CO\,(2-1) contours integrated over the 235--240\,km\,s$^{-1}$, 240--245\,km\,s$^{-1}$, 245--250\,km\,s$^{-1}$ are shown in magenta, green and red respectively and given in the bottom right corner, with the peak velocities of each cloud. The scale bar marks 10$''$ (at the distance to 30 Doradus, 10$'' \approx$ 2.5\,pc). (b) The total integrated CO\,(2-1) emission over entire velocity range with the colour scale giving the intensity of the emission in 10$^2$K\,km\,s$^{-1}$. The CO contours from Fig.\,2a, and the region delineating the stapler nebula is shown in black dashed lines. (c) Greyscale VLT-ISAAC image of H$_2$\,2.12$\mu$m emission showing the excited molecular gas in the stapler nebula. Stretch is in logarithmic scale. The corresponding cold molecular gas from Fig.\,2a, and the boundary of the stapler region is shown. The clumpy H$_2$ emission coincides spatially with the centres of each CO cloud knot. (d)  Greyscale Br$\gamma$ emission, with the contours from Fig.\,2a, and the boundary of the stapler nebula overlaid. Stretch is in linear scale. In contrast to the H$_2$ emission, the Br$\gamma$ emission shows no strong spatial coincidence with the centres, but rather with the weakest CO\,(2-1) contours. A description of the data presented in the figure is found in Section 2, with a discussion in Section 4.}
\end{figure*}
 \label{fig:COa}


\section{Data}

\subsection{{\rm CO\,(2-1)} observations}


We conducted a deep CO\,(2-1) survey centred on the 30 Doradus Nebula using the Swedish-ESO submillimeter telescope (SEST) across March 1997- January 2001. SEST was a 15\,m radio telescope located at La Silla, Chile.  The angular resolution at the CO\,(2-1) frequency of 230\,GHz is 23$\arcsec$, which corresponds to a projected size of 5.6\,pc at the distance to the LMC. 


Observations were conducted in position switching mode using a reference point free of CO\,(2-1) (at $\alpha$\,=\,05$^{h}$37$^{m}$54$^s$, $\delta$\,=\,$-69^{\rm \circ}$04$'24''$) for sky subtraction. 
The backend narrow-band high-resolution spectrometer (HRS) was used with a bandwidth of 80\,MHz, and a frequency resolution of 41.7\,kHz, which translates to a velocity resolution of 0.054\,km\,s$^{-1}$ at the frequency of CO\,(2-1). The data were reduced using the GILDAS software \footnote{http://www.iram.fr/IRAMFR/GILDAS}, with linear or third-order polynomial for baseline fitting. The resultant spectra were smoothed to a velocity resolution of 0.25\,km\,s$^{-1}$. The rms noise achieved in a single channel is 0.07\,K after 240\,s of integration. We mapped the 30 Doradus region with 10$\arcsec$ spacings and detected CO\,(2-1) emission across 30 Doradus, including the stapler nebula where CO\,(1-0) emission had not been previously detected (Pineda et al. 2009; Johansson et al. 1998).  In Fig.\,1, the CO\,(2-1) contours are overlaid on a {\it Hubble space telescope} (HST) three colour optical $BVI$H$\alpha$ image of 30 Doradus, where the position of CO\,(2-1) emission with respect to the R\,136 cluster, and its ionized surroundings can be visualized. Our observations represent a five-fold increase in sensitivity at twice the spatial resolution of previous CO\,(1-0) observations across the 30 Doradus nebula (see Pineda et al. 2009). Higher angular resolution observations of the 30 Doradus nebula are presented in Indebetouw et al. (2013), Anderson et al. (2014) and Nayak et al. (2016), but those data do not cover the region studied here.




\subsection{Near-infrared emission line imaging}

We obtained nIR imaging of the 30 Doradus region in H$_2$ 2.12$\mu$m narrowband filter, and the $K$s broadband filter using the ISAAC (Infrared spectrometer and Array Camera) imager mounted on 8m Melipal (UT3) telescope of the Very Large Telescope (VLT) situated in Paranal, Chile (program ID 078.C-0487A). Toward the R\,136 region, the mosaic covered a field of 5$\arcmin$$\times$5$\arcmin$ area. The average seeing measured from the images is $\sim$0.8$''$--1.1$''$. Observations were taken in the ABBA sequence with the sky image 30$'$ from the stapler nebula at $\alpha$\,=\,05$^h$39$^m$01$^s$, $\delta\,=\,-$69$^{\rm \circ}$42$'$36$''$ in a region free of nebulosity within the ISAAC field of view, ensuring adequate sky subtraction. The total integration time on source was 1 hour for the narrowband filter, and 100s for the $K$s broadband filter. Data were reduced using the ISAAC pipelines, with flux calibration carried out using Persson (1998) standards. Astrometric calibration was refined using 2MASS. Bright stars are saturated in the emission line images and have a non-linear CCD response, meaning they cannot be completely subtracted. This leads to circular bright residuals (and in some cases vertical bleeding) in the emission line images. We excluded these regions from further analysis by masking them. 

We used the Br$\gamma$ 2.165$\mu$m narrowband flux calibrated image from Yeh et al. (2015). The image was obtained using the NOAO Extremely Wide Field Infrared Imager (NEWFIRM) mounted on the 4m Victor Blanco telescope located at the Cerro Tololo Inter-American Observatory, Chile. The final Gaussian convolved image resolution for the Br$\gamma$ narrowband image is 1$\arcsec$, and is comparable to the VLT H$_2$\,2.12$\mu$m narrowband image. 

\subsection{Archival photometry}

mIR photometry of point sources in the stapler nebula were estimated by Gruendl \& Chu (2009) from images at 3.6, 4.5, 5.8 and 8.0$\mu$m taken using the {\it Spitzer} space telescope IRAC (Infrared Array Camera) as part of the {\it Spitzer} legacy program SAGE (Spitzer Survey of the Large Magellanic Cloud: Surveying the Agents of a Galaxy's Evolution; Meixner et al. 2006). The full width half maximum (FWHM) of the images is 1.6$\arcsec$, 1.7$\arcsec$, 1.7$\arcsec$ and 2$\arcsec$ respectively. Alternative photometry of point sources from the same images are also presented by Whitney et al. (2008), but Whitney et al. (2008) miss a significant fraction of point sources in the 30 Doradus region, as their study is motivated towards detecting reliable sources throughout the LMC via pipeline analysis. Gruendl \& Chu (2009) detect objects in the dense and nebulous surroundings using detailed aperture photometry (see Section 6.3 of Gruendl \& Chu (2009) for a comparison). Photometry in the fIR at 100 and 160$\mu$m, and 250 and 350$\mu$m of point-like and extended sources in the stapler nebula is given in Seale et al (2014), using images taken by the {\it Herschel} space telescope PACS (Photoconductor Array Camera and Spectrometer), and SPIRE (Spectral and Photometric Imaging Receiver) instruments respectively as part of the {\it Hershel} large program HERITAGE (HERschel Inventory of The Agents of Galaxy Evolution; Meixner et al. 2013). The FWHM for these images are 8$\arcsec$, 12$\arcsec$, 18$\arcsec$ and 25$\arcsec$ respectively. We utilize the photometry from Gruendl \& Chu (2009) and Seale et al. (2014) in this study.

\section{Molecular clouds}

Figure \ref{fig:CO} shows the CO\,(2-1) integrated line emission over the velocity interval 235--270 \,km\,s$^{-1}$ as contours, superimposed on the HST composite $BVI$H$\alpha$ image. Strong CO\,(2-1) emission from Cloud 10 (northeast region of the map) and Cloud 6 (southwest region of the map) previously reported by Johansson et al. (1998) is seen along the northeast-southwest axis.

\begin{figure}
\includegraphics[width=0.495\textwidth]{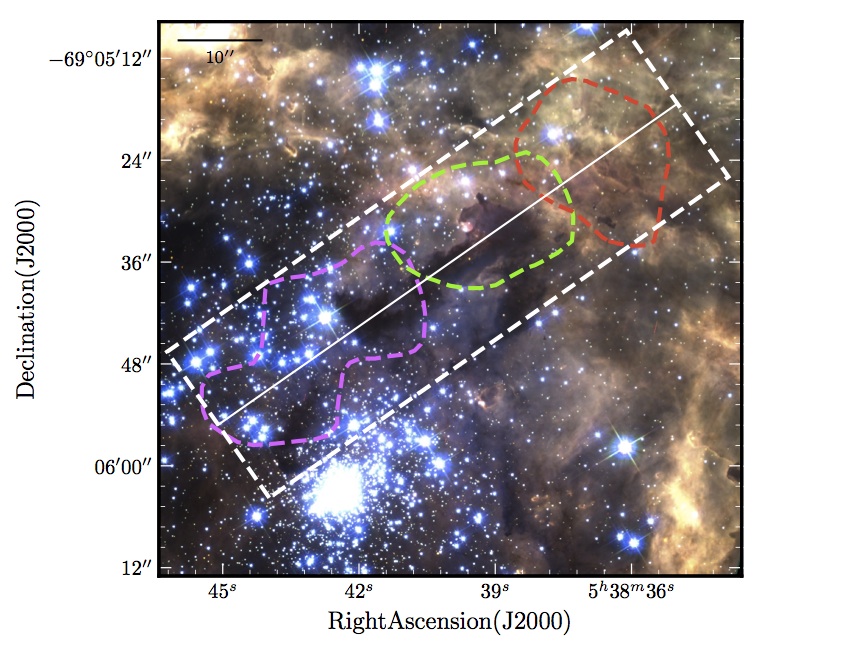}
\includegraphics[width=0.495\textwidth]{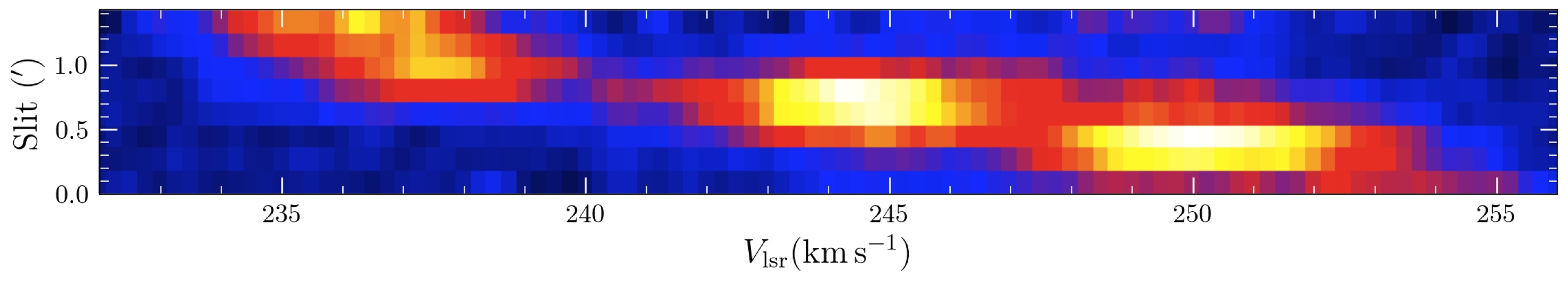}
\includegraphics[width=0.495\textwidth]{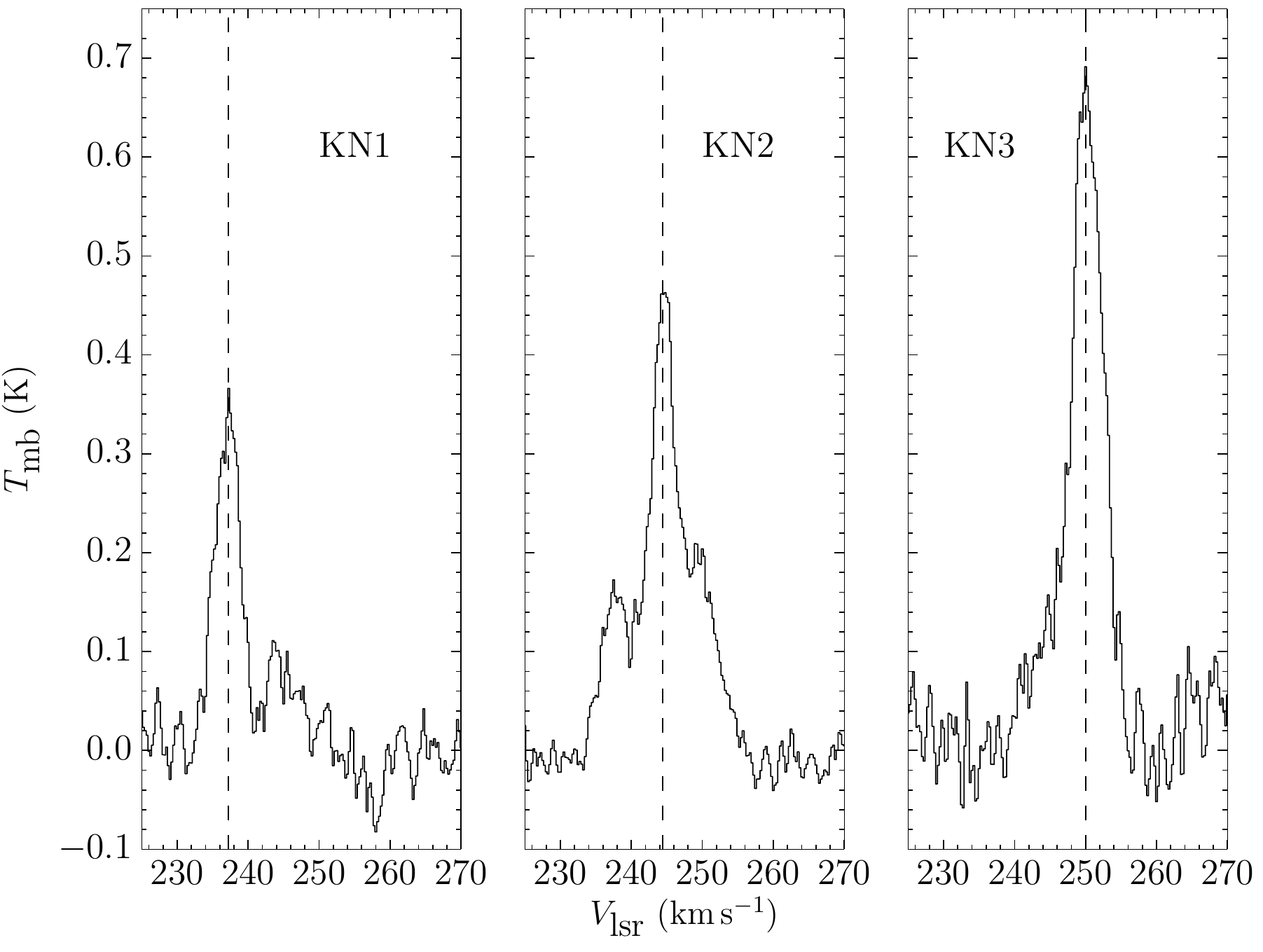}
\caption{{\it Top}: The stapler nebula blown up from the HST mosaic in Fig.\,1, with the stapler region is outlined with the dashed white box. The outermost CO\,(2-1) contours integrated over the 235--240\,km\,s$^{-1}$, and 245--250\,km\,s$^{-1}$ are shown in magenta and red respectively, with the second outermost contour of the 240--245\,km\,s$^{-1}$ also shown. The solid white line marks the position of the slice shown in the middle panel. {\it Middle}: Position velocity slice of the CO\,(2-1) cube in linear scale along the direction of the slit given by the solid white line in the top panel. The slit cuts along the centre of the stapler nebula. {\it Bottom}: CO\,(2-1) spectra of each cloud extracted from the region bounded by the contours shown in top panel. The dashed lines for each cloud is it's $V_{\rm{lsr}}$ given in Table 1. }
    \label{fig:spectra}
\end{figure}

We observe previously undetected CO\,(2-1) emission originating from the region located between Clouds 6 and 10, close to R\,136 (see Fig.\,\ref{fig:COa}). The emission extends along the southeast-northwest direction in projection. The distance of the emission from R\,136 in projection is between 2\,pc (emission is located to the north of R\,136) to 14\,pc away (in the northwest direction from R\,136). This CO emission is approximately five times weaker than the CO emission observed in Clouds 6 and 10. This emission is coincident spatially to the stapler nebula in the optical image of Fig.\,\ref{fig:COa}a. The emission is resolved in CO\,(2-1) velocity as a chain of small and weak clouds (Fig.\,\ref{fig:COa}a,b). We define the stapler region by the extent of the CO\,(2-1) emission, which goes beyond the visible stapler shaped dark cloud in the optical. This boundary is marked in Fig.\,1 with a dashed rectangle. We named the CO clouds Knots (KN), as they form a chain separated in velocity, as demonstrated by the position velocity slice across the stapler nebula, and the CO\,(2-1) spectra of each individual cloud shown in Fig.\,\ref{fig:spectra}. By analysing the radial velocities and spatial distribution we found that the KN clouds are composed of three clouds we name KN1, KN2, and KN3 in order of decreasing Right Ascension (labelled in Fig.\,\ref{fig:COa}a).

\subsection{Physical properties}

After identifying each cloud, we computed the central velocity ($V_{\rm lsr}$) in the local standard of rest frame, and velocity width ($\sigma_{v}$) by fitting a Gaussian profile to the total cloud spectrum. The major and minor axis sizes of the profiles, in conjunction with the rms size of the beam were used to compute the deconvolved radius ($r$). Given the uncertainties on the Gaussian fit of the CO spectra found for each cloud are around 30\%, we estimate the uncertainties on $r$ to be 15\%.  

\subsubsection{CO luminosity and mass}

The CO cloud luminosity is computed as:
\begin{equation}
L_{\rm CO}\, [K{\rm kms}^{-1}{\rm pc}^{-2}] = \rm{D}^2 \int_{\Omega} \int_v T_{\rm mb}(\nu) \, d\nu \,d\Omega
\end{equation}
where D is the distance to the source in pc (adopted as 50\,kpc), $T_{\rm mb}$ the main beam
temperature which is the antenna temperature corrected for the efficiency
of the antenna ($T_{\rm mb} = T_{\textrm{A}}/\eta$), and $\Omega$
is the solid angle of the subtended by the source.   

The H$_2$ mass of the clouds can be calculated from the observed
CO\,(1-0) luminosity assuming a linear conversion between the velocity
integrated CO emission ($I_{\textrm{CO}}$) and the H$_2$ column density ($N_{{\rm H}_2}$);
\begin{equation}
N_{\rm H{_2}} = X_{\textrm{CO}}\, [cm^{-2}(K{\rm kms^{-1}})^{-1}]\,\, I_{\textrm{CO}}\,[K{\rm kms^{-1}}],
\end{equation}
where $X_{\textrm{CO}}$ is the CO-to-H$_2$ conversion factor (Bolatto et al. 2013; Roman-Duval et al. 2014). The total mass of H$_2$ ($M_{{\rm H}_2}$) is,  
\begin{equation}
  M_{\textrm{H}_2} \,[M_\odot] = \alpha_{\textrm{CO}} {\rm D}^2\, [{\rm Mpc}] S_{\textrm{CO}}, 
\end{equation}
where 
\begin{equation}
  \alpha_{\textrm{CO}}\, [M_\odot {\rm Mpc}^{-2} ({\it Jy}{\rm kms}^{-1})^{-1}] = X_{\textrm{CO}} \frac{m_{\textrm{H}_2} c^2}{2 k \nu^2},
\end{equation}
and the flux density $S_{\textrm{CO}}$ is, 
\begin{equation}
S_{\textrm{CO}}\,\, [{\it Jy}{\rm kms}^{-1}] = \int S_{\nu} \, dv.
\end{equation}
The molecular gas mass is multiplied by 1.36 to include the Helium contribution. This method is calibrated for the $J$=1$\rightarrow$0 transition. We use a ratio between the CO\,$J$=2$\rightarrow$1 and $J$=1$\rightarrow$0 lines of 0.87 for 30 Doradus Cloud 10 (the North Eastern cloud; see Fig.\,1) found by Johannson et al. (1998) to estimate the CO\,(1-0) luminosity.

The conversion factor depends on both metallicity and the ambient radiation field intensity (Maloney 1988). As a consequence of strong radiation fields and poor self-shielding in low metallicity
environments, the CO molecule is photo-dissociated as it does not self-shield like the H$_2$ molecule. Therefore, there is less CO
compared to the H$_2$ abundance in the Magellanic Clouds than in the
Galaxy. This translates into higher values of $X_{\textrm{CO}}$ in the LMC compared to the Galaxy. In general the conversion factor increases
with higher radiation fields and decreases with higher metallicities.  We adopt the median conversion factor in the LMC compiled from the literature by Bolatto et al. (2013) of
\begin{equation}
  X_{\textrm{CO}} = 8.8 \pm 0.3 \times 10^{20} \,[\textrm{cm}^{-2} \textrm{(K kms}^{-1})^{-1}].
\end{equation}
The adopted $X_{\textrm{CO}}$ factor is 3.8 times larger than the canonical Galactic $X_{\textrm{CO}}$ of Bolatto et al. (2013). Our adopted value is similar to that found by Herrera et al. (2013) when comparing molecular and dust mass estimates in the LMC N11 region; but higher than the value of $6 \times 10^{20} \,\textrm{cm}^{-2} \textrm{(K kms}^{-1})^{-1}$ reported by Roman-Duval et al. (2014). The resulting cloud masses if we adopted the Roman-Duval et al. (2014) $X_{\textrm{CO}}$ would be reduced by $\sim20$\%.

\subsubsection{Virial mass}
The virial mass ($M_{\rm{vir}}$) was computed assuming that each cloud is spherical, is in
virial equilibrium and has a density ($\rho$) profile of the form $\rho \propto
r^{-1}$. The virial mass is given by
\begin{equation}
  M_{\textrm{vir}} \,[M_\odot] = 190 \sigma_{v}^2\,[{\rm kms}^{-1}]\, r\, [{\rm pc}]
  \label{eq:virial_mass2}
\end{equation}
according to MacLaren et al. (1988). The results of our analysis for each cloud are given in Table 1.

\subsection{Analysis}

\subsubsection{Larson's Laws}

\begin{figure*}
		\plottwo{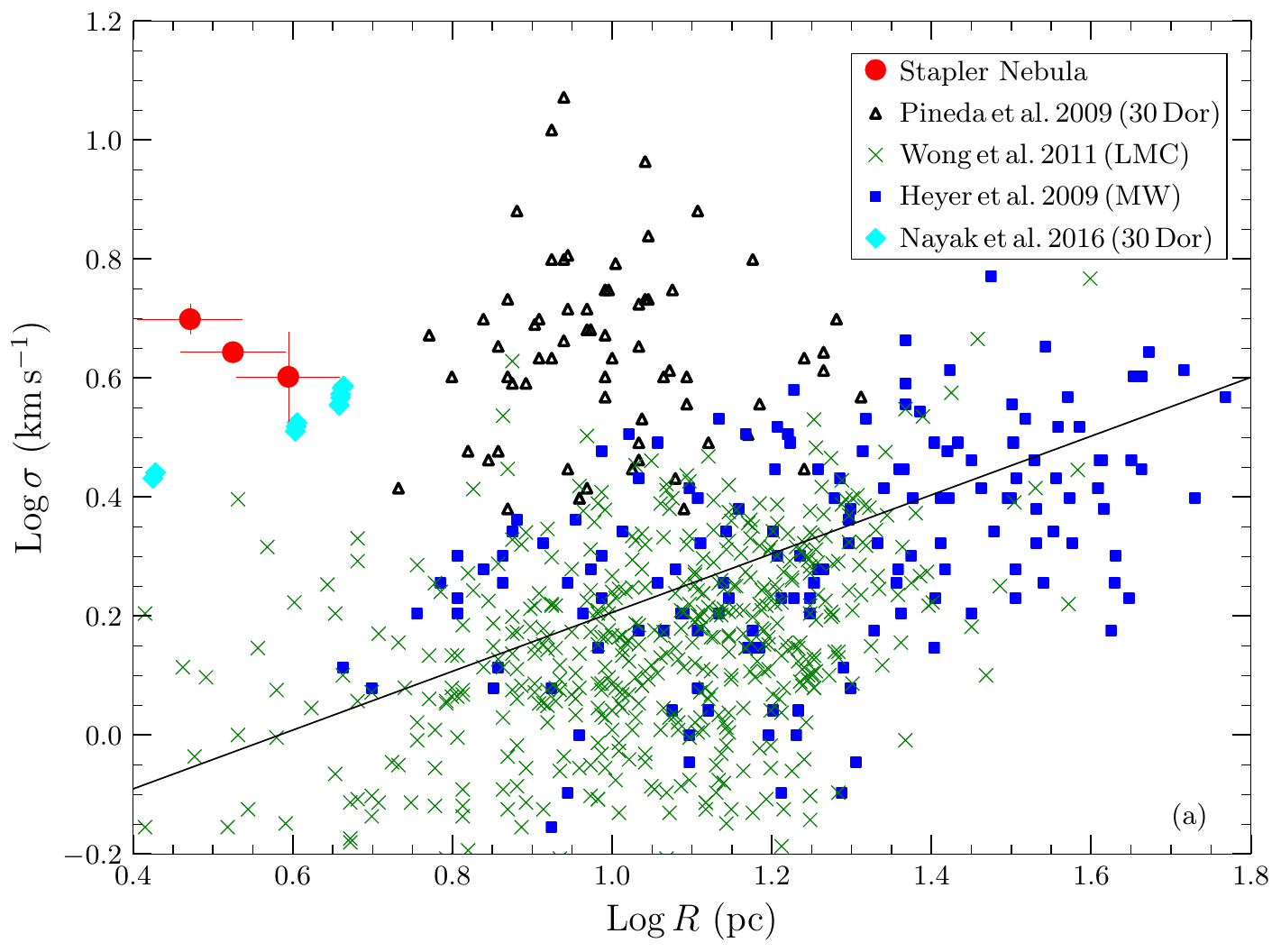}{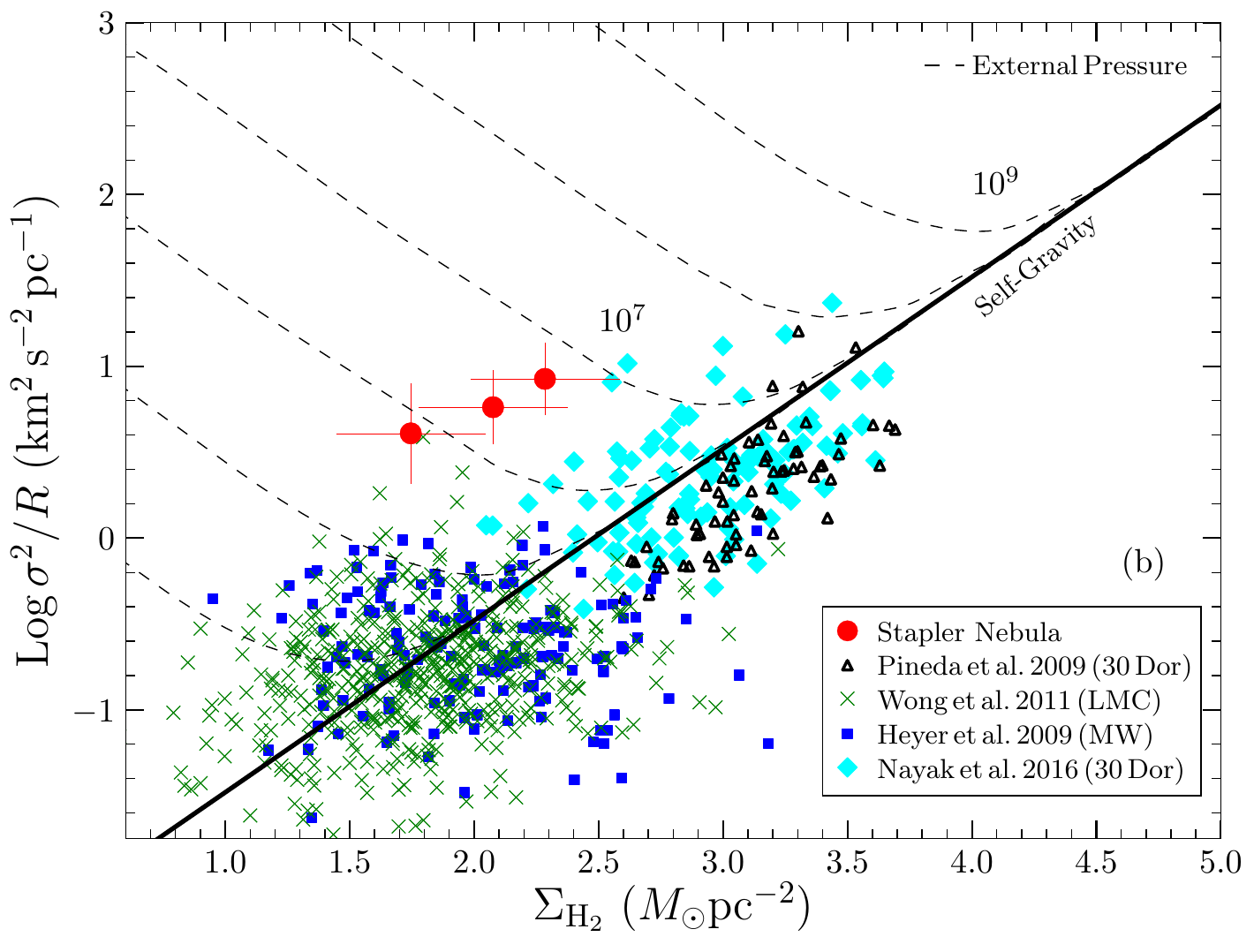}
    \caption{(a) Linewidth size relation for CO\,(2-1) clouds identified near R\,136 (circles). The positions of clouds in the 30 Doradus region from Pineda et al. (2009) are shown as triangles, and green crosses mark clouds identified in the LMC by Wong et al. (2011). Blue squares are Galactic molecular clouds from Heyer et al. (2009), which follow the canonical relation of $\sigma_{v}$=0.72\,$R^{0.5}$. (b) The ${\sigma_v}^2/r$-$\Sigma_{\rm H_2}$ relation of the stapler molecular clouds, plotted alongside those from the literature (symbols same as Fig.\,4a). The solid line gives the approximate ${\sigma_v}^2/r$ value for increasing $\Sigma_{\rm H_2}$ of an isolated virialized cloud confined by self-gravity. The dashed lines marks the equilibria for external pressures between $P$/$k_{\rm B}$$\sim$10$^3$ to 10$^9$\,cm$^{-3}$\,K of a centrally concentrated cloud approximated by hydrostatic equilibrium. The LMC (incl. those in 30 Doradus) and Galactic molecular clouds appear confined by self-gravity, whereas the KN clouds require pressures $\gtrsim$10$^6$\,cm$^{-3}$\,K to confine them. The high line widths therefore reflect that the clouds are likely confined by external pressure.}
    \label{fig:slw}
\end{figure*}

Molecular clouds in virial equilibrium follow the empirical power law relation $\sigma_{v} \propto r^{\alpha}$ (Larson 1981). $\alpha$ is generally agreed to be between 0.4--0.5 based on numerous molecular cloud surveys of the Milky Way, and external normal and dwarf Galaxies (Bolatto et al. 2008; Heyer et al. 2009). The observed value of $\alpha$ is oft explained by turbulence (McKee \& Ostriker 2007; Lombardi et al. 2010). The velocity dispersion is considered to be a measure of the internal dynamics within the clouds, because the observed line profiles, averaged over a cloud, have Gaussian shapes and the line widths are broader than the thermal line widths. It is a reasonable assumption to make that the line profiles are produced by turbulent motions of the gas inside the clouds (Solomon et al. 1987).

Figure\,\ref{fig:slw}a displays the position of the KN clouds in the $\sigma_{v}$--$r$ diagram. Also shown are results from Heyer et al. (2009) summarising the canonical relation found for Galactic clouds of $\sigma_{v} = 0.72\,r^{0.5}$; and clouds in the 30 Doradus region (from Pineda et al. 2009; having a resolution of 43$\arcsec$; and from Nayak et al. (2016) having a resolution of 2$\arcsec$), and in the LMC (excluding 30 Doradus) from Wong et al. (2011) whose study had a spatial resolution of 45$\arcsec$. Note that the Pineda et al. (2009) and Nayak et al. (2016) clouds do not cover the central region of 30 Doradus near R\,136, and there are no spatial overlaps between the KN clouds and the clouds they identify. The molecular clouds associated with the stapler nebula lie above the canonical $\sigma_{v}$--$r$ relation for Galactic clouds. Interestingly, the other detected clouds in 30\,Doradus from Pineda et al. (2009) and Nayak et al. (2016) also lie above the canonical relation. Although the departure from the relation is not at the same scale as the KN clouds, this might still indicate that the observed $\sigma_{v}$--$r$ relation might be a function of distance from R\,136, and a more global property of 30\,Doradus.

The position of the KN molecular clouds in the $\sigma_{v}$--$r$ diagram implies either of two scenarios; the clouds are collapsing or expanding, or the observed line widths are the manifestation of external pressures that keep the clouds in equilibrium. Since collapse velocities are generally only $\sim40$\% larger than equilibrium velocity dispersions for a self-gravitating cloud, the observed large linewidths for the sizes of the KN clouds are probably not the result of collapse (Ballesteros-Paredes et al. 2011). Neither are they likely to result from expansion because there is no obvious shell or hole structure that usually accompanies expansion.

\begin{table*}
	\centering
	\caption{Properties of the molecular clouds.}
	\label{tab:example_table}
\begin{tabular}{lcccccccc}        
\hline\hline                 
Name & R.A. & Dec. & $V_{\rm lsr}$ & $\sigma_{\rm v}$ & $L_{\rm CO}$ & $r$ & $M_{\rm H_2}$ & $M_{\rm vir}$\\  
 & (J2000) & (J2000) & (km\,s$^{-1}$) &  (km\,s$^{-1}$) & (\,K\,km\,s$^{-1}$\,pc$^2$)  & (pc)  & (10$^3\,M_{\odot}$) & (10$^3\,M_{\odot}$)\\    
\hline
    30Dor-KN1 & 5$^h$38$^m$45$^s$ & $-$69$^{\rm \circ}$06$'$00$''$ & 237.2$\pm$0.1 & 4.0$\pm$0.7 & 154.2$\pm$21.3  & 3.93$\pm$0.58 & 2.7$\pm$0.2 & 10.9$\pm$3.3  \\
    30Dor-KN2 & 5$^h$38$^m$40$^s$ & $-$69$^{\rm \circ}$05$'$30$''$ & 244.4$\pm$0.2 & 4.4$\pm$0.1 & 243.6$\pm$3.7  & 3.35$\pm$0.5 & 4.2$\pm$0.3 & 11.25$\pm$1.8  \\
    30Dor-KN3 & 5$^h$38$^m$36$^s$ & $-$69$^{\rm \circ}$05$'$30$''$ & 250.0$\pm$0.2 & 5.0$\pm$0.3 & 320.8$\pm$44.3  & 2.96$\pm$0.44 & 5.3$\pm$0.4 & 12.83$\pm$2.5  \\
\hline                        
      
\end{tabular}\\
\end{table*}

We examine whether the observed large $\sigma_{v}$ are the manifestation of external pressures necessary to keep the clouds in equilibrium. In Fig.\,\ref{fig:slw}b, we plot the mass surface density ($\Sigma_{\rm H_2}$) against the ${\sigma_v}^2/r$ value of the KN clouds, along with those in the Milky Way from Heyer et al. (2009); in 30 Dor from Pineda et al. (2009) and Nayak et al. (2016); and in the LMC from Wong et al. (2011). Isolated virial clouds confined by self-gravity follow a linear relation in the ${\sigma_v}^2/r$ vs. $\Sigma_{\rm H_2}$ plot (for e.g. see Heyer et al. 2009). The values of clouds from the literature fall along this expectation. However, the KN clouds alone depart from the expected relation, and are likely confined by external pressure and not in virial equilibrium. Following the simplifying assumptions of Field et al. (2011), we plot isobars of external pressure (in terms of $P/k_{\rm B}$) according to their prescription. The lines reflect the external pressure necessary to confine clouds for a given ${\sigma_v}^2/r$ assuming clouds with a centrally concentrated internal density structure approximated by hydrostatic equilibrium. From Fig.\,\ref{fig:slw}b, we see that external pressures of $\sim\,10^6$\,cm$^{-3}$K are necessary to keep the KN clouds confined. These values are in general agreement with Chevance et al. (2016), who report that the stapler nebula is located in a region with gas pressure $\sim$ 0.85-1.2$\times 10^6$\,cm$^{-3}$K, with the peak found in KN2 (see Fig. 15 in Chevance et al. 2016).

\begin{figure*}
	\plottwo{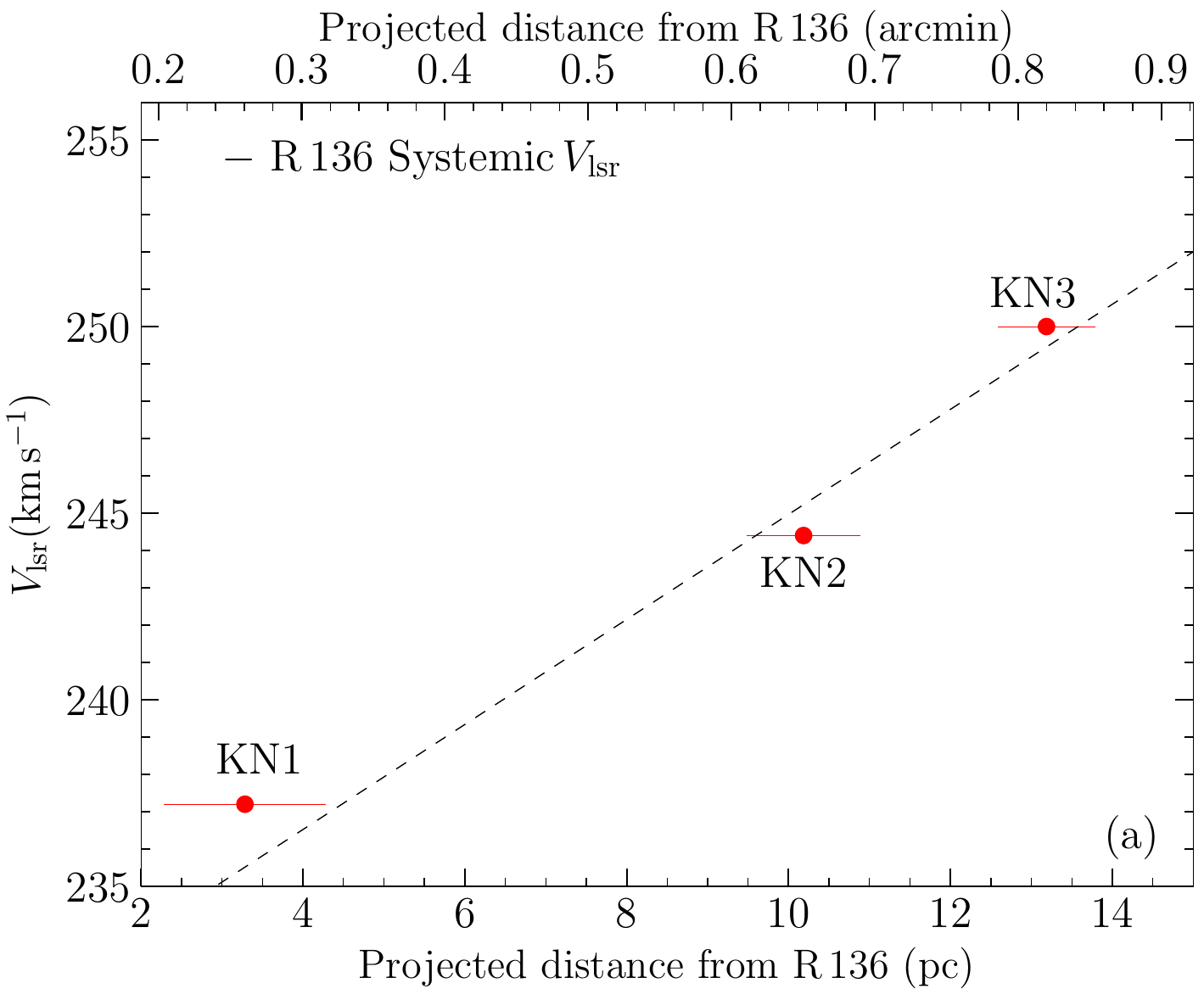}{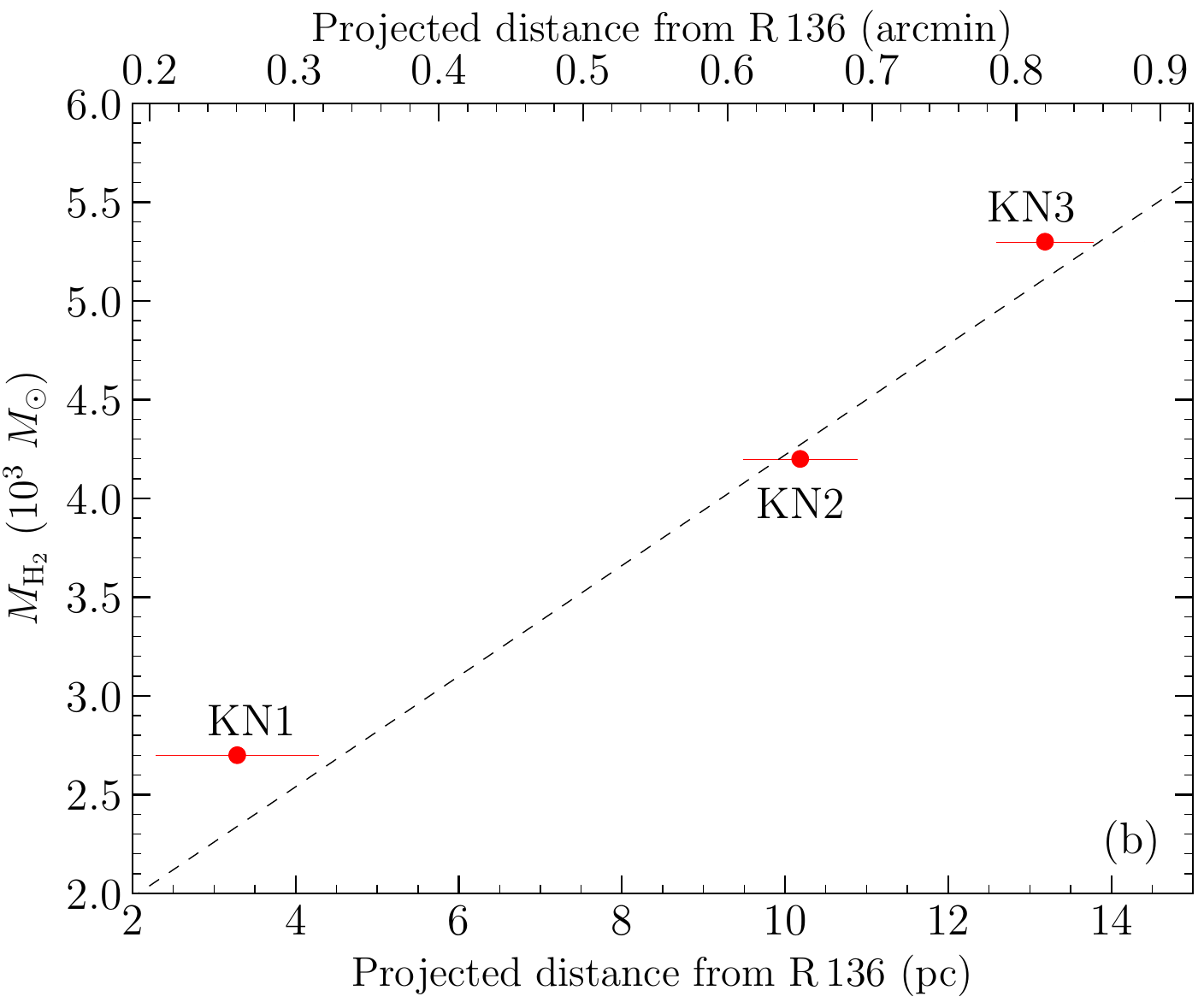}
    \caption{(a) The peak velocity in the local standard ($V_{\rm lsr}$) of rest frame of each cloud, is plotted as a function of projected distance from R\,136. The upper y-axis gives the projected distance in arcmin, while the lower y-axis scales in parsecs assuming a distance of 50\,kpc. The best fit relation is given by the dashed line, and has a slope of $\sim$1.5. (b) The mass of each cloud $M_{\rm H_{2}}$, is plotted as a function of projected distance from R\,136. The upper y-axis gives the projected distance in arcmin, while the lower y-axis scales in parsecs assuming a distance of 50\,kpc. The best fit relation is given by the dashed line.}
\label{fig:distanced}
\end{figure*}

\subsubsection{Variation in properties as a function of distance from R\,136}

The $V_{\rm lsr}$, and $M_{{\rm H}_2}$ of each cloud are plotted as a function of projected distance from R\,136 in Fig.\,\ref{fig:distanced}. The velocity of the clouds increases as a function of projected distance from R\,136. The velocity of the KN3 cloud loosely matches the radial velocities of the stars within R\,136\footnote{The mean radial velocity of the stars translated to local standard of rest frame is $\approx$255$\pm$5\,km\,s$^{-1}$ (Evans et al. 2015)}. The KN1 cloud is closest in projection to R\,136 and is blue shifted with respect to the mean velocity of stars in the cluster. This suggests the KN1 cloud, (and likely KN2 and KN3 clouds as suggested by the lack of background stars) lie slightly in front of the cluster. The clouds appear to be moving away from the cluster as function of projected distance from it (Fig.\,\ref{fig:distanced}a).

The $M_{{\rm H}_2}$ of each cloud also increases as the projected distance from R\,136 increases (Fig.\,\ref{fig:distanced}b). The KN1 cloud mass is approximately 2 times lower than the KN3 cloud. Under the assumption that the molecular clouds detected in CO\,(2-1) were initially all of similar densities, and photoionization from R\,136 alone is evaporating the molecular cloud, then KN1 must be closer to R\,136 (considering it as the only source of external photoionization) because the ionizing flux decreases to the inverse square with distance.  



\subsection{Line of sight distances}

Our main results concerning the detection of cold molecular gas near R\,136 suffers from possible projection effects.
Although the cold molecular gas detected in the CO\,(2-1) observations lies within 2-14\,pc in projection of the R\,136 cluster, the actual distance may likely be further in the line of sight direction allowing for the clouds to possibly survive photoionisation. Chevance et al. (2016) analyse the physical distance of the CO gas in 30 Doradus to the stars, by comparing the incident radiation field on the gas modelled against fIR observations in fine structure lines of the emitted radiation field measured from the known massive star population. By comparing the luminosity of the photodissociation region (which forms the interface between the photoionizing radiation from the stars and the gas) against their predictions, they are able to constrain the line of sight distance of the photodissociation regions from R\,136 with uncertainties of 4\,pc. Based on their results (Figure 20 in Chevance et al. 2016), the stapler nebula lies less than 20\,pc away in the line of sight direction from R\,136, which itself lies at the centre of a sphere of about 6\,pc in radius. This distance agrees well with the line of sight distance measured from line ratios of ionized lines in optical spectra by Pellegrini et al. (2010). From their Fig.\,12, we find that the distance of the KN clouds is less than 20\,pc away in our line of sight from R\,136.

We also consider that if the CO gas is close to R\,136, and coincident with the dust, there is likely to be a gradient (reflecting the gradient in the projected $V_{\rm lsr}$) in the dust temperature, and the total fIR luminosity arising from the photodissociation region. Such a gradient is visible in both the dust temperature maps (Guzman 2010), and also in the total fIR luminosity which peaks at KN2, and decreases towards KN3 (see Fig.\,1 of Chevance of et al. 2016). Therefore, although our observations are unable to resolve the line of sight distances to the KN clouds from R\,136, based on corroboration from multiple independent sources in the literature, we find that the line of sight distance to the KN clouds from R\,136 is $\lesssim$20\,pc. The molecular clouds in the stapler nebula lie between 2-14\,pc in projected distance, and $\lesssim$20\,pc in the line of sight distance from R\,136.

\section{Near-infrared emission line imaging}

The H$_2$ 2.12$\mu$m emission line image is shown in Fig.\,\ref{fig:COa}c, with the CO\,(2-1) contours overlaid. Strong H$_2$ emission is spatially coincident with the CO\,(2-1) emission of the molecular clouds, and shares similar morphology. The H$_2$ emission is clumpy, with numerous knots and a reticulated pattern. In contrast, detected ionized gas (Br$\gamma$) at the position of the CO\,(2-1) molecular clouds is weak and diffuse (see Fig.\,\ref{fig:COa}d). This diffuse Br$\gamma$ emission is associated with filaments and arc-like structures of ionized gas vivid in H$\alpha$ (brown in Fig.\,\ref{fig:COa}a). This convinces us that the strong H$_2$ nIR emission is the warmer component of the cold molecular gas traced by the CO detections, whereas the diffuse Br$\gamma$ emission lies slightly beyond the ionized surface of the molecular cloud although no clear demarcation is noted. The Br$\gamma$ morphology is not spatially coincident with the H$_2$ and CO\,(2-1) emission. 


The H$_2$/Br$\gamma$ ratio can be used to disentangle shock/collisionally excited H$_2$ from fluorescence excitation. This is because the shocks and collisional excitation affect primarily the H$_2$ gas, leaving the ratio of H$_2$/Br$\gamma$ above unity, whereas fluorescence acts on both the molecular and ionised gas leading to a ratio below unity. Using the absolute flux ratio, we find that the H$_2$/Br$\gamma$ ratio never exceeds 0.5 at an angular resolution of 1$\arcsec$ in the stapler nebula, agreeing with the findings of Yeh et al. (2015). This indicates that the excited nIR H$_2$ is primarily excited by the ultraviolet (UV) radiation from R\,136 acting on the surfaces of the molecular gas, with the filamentary Br$\gamma$ arising from the same source. The KN clouds therefore must lie in front of us given the morphology of the clumped H$_2$ emission, and lack of background stars. We note that it is possible that shock excited emission is prevalent on smaller scales (a few tenths of a parsec, or $\lesssim$0.5$\arcsec$ at the distance to 30 Doradus) caused by outflows from massive protostars residing within the KN molecular clouds, but our current angular resolution limitations in the nIR narrowband images ($\sim$1\,$\arcsec$) prevent us from examining the same in detail. Future high angular resolution integral field unit (IFU) nIR spectroscopy would help towards constraining this further, as shocked gas will likely be offset in velocity.

A picture of ionization fronts emerges, with the photodissociation region extincted from our line of sight by the cold molecular gas. These structures could resemble the dense ``pillars'' or structures observed in the galaxy in regions such as M16 and NGC\,3603 (Sankrit \& Hester 2000), but are smaller in scale at $\sim$0.1--0.3\,pc, and in the 30 Doradus Nebula by Pellegrini et al. (2010). From the observed emission line imaging and CO\,(2-1) observations it appears that the clouds are being ionised on the backside. We are viewing the KN molecular clouds face on, and they are likely the tail of pillar-like structures (we refer the reader to Pound 1998 for a description of pillar morphology; or to Fig.\,7) with the ionized head pointing towards R\,136. The observed velocity line widths of the CO\,(2-1) line are then likely caused due to the velocity gradient between the head and tail of pillars (e.g. Pound 1998). Following the Bertoldi (1989) analytical theory of photoevaporating clouds, during photoevaporation clouds form a cometary structure similar to pillars. Neutral gas at the head is pushed back by the ionization front stripping the outer envelope. The difference in velocity between the slowly moving head to the tail with respect to ionizing stars develops a velocity gradient, that is manifested as velocity line width seen in Fig.\,\ref{fig:slw}, also providing a natural explanation for the observed molecular cloud properties. The observed velocity line widths is in a similar range to those observed in M16 using CO observations (Pound 1998). Our results are well supported by the work presented in Chevance et al. (2016). The stapler nebula is prominent in forbidden mIR and fIR line emission from [S{\scriptsize II}], [C{\scriptsize II}], and [O{\scriptsize I}] (Fig. 1 in Chevance et al. 2016), which are key tracers of photodissociation regions.

\section{Candidate massive young stellar objects}

\subsection{Identification of YSO}
\begin{figure*}
	\includegraphics[width=2.1\columnwidth]{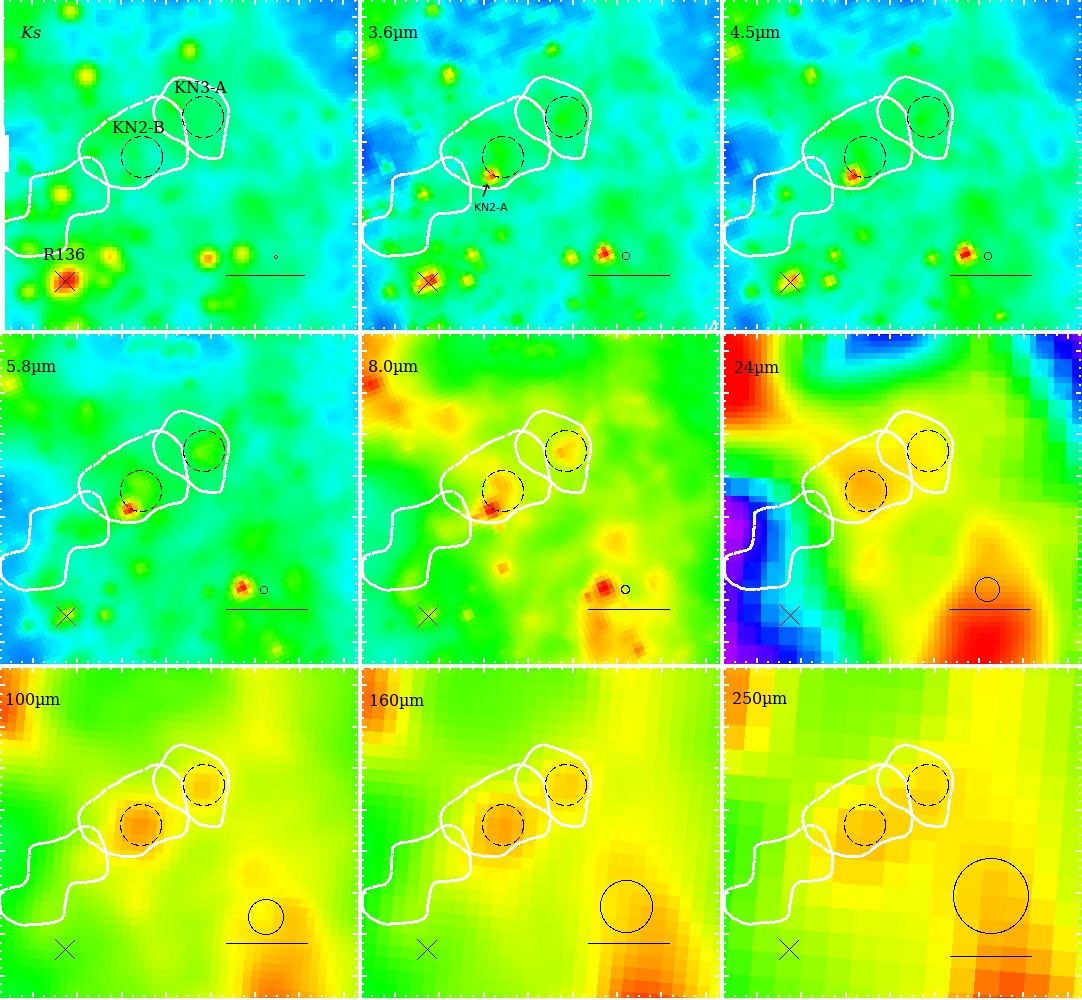}
    \caption{$K$s-250$\mu$m inverted colour thumbnails with the positions of KN2-B (bottom) and KN3-A  (top) overlaid as dashed circles. The stretch is in logarithmic scale. Each circle has a radius of 5$''$, while the cross marks the position of R\,136. The scalebar in the bottom right corner is of 20$''$, or $\approx$5 pc at the distance to the LMC. The FWHM of each image is given by the solid circle above the scalebar. From the images, KN2-B and KN3-A are not visible at short wavelengths, but have PAH emission associated with them in the 8$\mu$m image (the PAH feature is at 8.6$\mu$m, and falls within the bandwidth). At 24$\mu$m, emission likely arising from protostellar heating of the surroundings is visible at the position of KN2-B, but not in KN3-A. At longer wavelengths  of 100 and 160$\mu$m from the {\it Herschel} imaging, both KN2-B and KN3-A are visible as point-like sources. However, at 250$\mu$m, KN2-B only appears significant, with KN3-A extended. Also shown are the outermost contours for each of the KN clouds from Fig.2a. In the 3.6$\mu$m, the position of KN2-A is given by the arrow. The source is not bright at $K$s or wavelengths $>$10$\mu$m.}
\label{fig:ysothumb}
\end{figure*}

Our aim is to detect any on-going star formation through the identification of YSOs using archival infrared photometry. The youngest YSOs (Class 0 objects) will only be visible at $\lambda>$10$\mu$m, while older Class I/II sources will be visible at shorter ($>$2\,$\mu$m) wavelengths. Using {\it Spitzer} 3.6-24$\mu$m, and {\it Herschel} 100-350$\mu$m photometry (details in Section 2.3; and images are presented in Fig.\,\ref{fig:ysothumb}), we classify point or point-like sources based on either the spectral energy distribution (SED) slope in the 3.6-24$\mu$m wavelength range, or by examining the parameters derived from fitting modified blackbody and YSO models across the entire wavelength range. We require detection in at least three bands in each catalogue to classify the object as a source. We then check for any counterparts with the coordinates of the detection in the {\it Herschel} 100$\mu$m within a crossmatch radius of 6.7$''$ (which is the convolution kernel at 100$\mu$m) and the {\it Spitzer} coordinates. We find a known {\it Spitzer} source in KN2 (KN2-A, or S5 in Walborn et al. 2013), a {\it Herschel} source without a {\it Spitzer} counterpart in KN2 visible at 100-350$\mu$m (KN2-B), and one {\it Herschel} source with a {\it Spitzer} counterpart within 0.9$''$ in KN3 (KN3-A). The thumbnails of these sources in the {\it Spitzer} and {\it Herschel} bands are shown in Fig.\,\ref{fig:ysothumb}.

For SED fitting, we utilize three methods to derive the properties of the source. First, we fit the photometry of each source with the YSO models of Robitalle (2017). The model grid covers 20000 radiative transfer YSO models covering a mass range of 0.1--50$M_{\odot}$. It should be noted that the SED fits are not intended to provide accurate parameters, but as a crude guide to the nature of each sources and that these models are limited in nature compared to the available free parameters (Offner et al. 2012). Secondly, we use a modified blackbody fit to the far-IR {\it Herschel} sources to constrain the temperature of the emitting source, $T_{\rm b}$. The observed flux can be reproduced as a blackbody with frequency flux density $F_{\nu}$ as
\begin{equation}
F_{\nu} =B_{\nu}(T_{\rm b})(1-e{^{-\tau_{\nu}}})\Omega.
\end{equation}

Here, $B_{\nu}(T_{\rm b})$ is the blackbody emission at $T_{\rm b}$, and $\tau_{\nu}$ is the optical depth constrained by a power law at frequency $\nu$ by $\tau_{\nu} \propto \nu^{\beta}$. $\beta$ = 1.5 in the LMC (Galliano et al. 2011). Thirdly, only for the {\it Spitzer} sources we employ the slope of the SED fit ($\alpha_{\rm SED}$), where
\begin{equation}
\alpha_{\rm{SED}}\,=\,\frac{d\,{\rm{log}}(\lambda F_{\lambda})}{d\,{\rm{log}}\lambda}.     
\end{equation} 
We consider only the flux between the wavelengths 3.6--24.0\,$\mu$m which classify well Class I/II sources (e.g. Greene et al. 1994). The results of our analysis are presented in Table 2, and described in the following subsections.

\subsection{Description of YSOs in the clouds}

\subsubsection{{\rm KN1} cloud}

KN1 contains no identified mid or far infrared source. 

\subsubsection{{\rm KN2} cloud}

KN2 is extremely promising as a future site of high mass star formation. KN2-B is a {\it Herschel} source detected in 100--350$\mu$m imaging (marked in Fig.\,\ref{fig:ysothumb}). There exists no near counterpart mIR source. The nearest mIR source is KN2-A which lies to its immediate south west, and at the very edge of the CO emission. This source angularly coincides with the dense region observed in CS to the south of KN2 by Rubio et al. (2009). CS is a tracer of dense gas with densities $n$, upwards of $10^5$\,cm$^{-3}$. 

KN2-B is classified as a high-mass YSO based on the {\it Herschel} classification scheme devised by Seale et al. (2014) for YSOs in the Magellanic Clouds. Following their three classification criteria (see Sec. 4.4 of that paper for further details), we find KN2-B meets all three as it is (i) the dominant source of fIR photometry and is clearly defined in at least 3 {\it Herschel} bands at 3$\sigma$ above the background (ii) It is not identified as a background galaxy/interloper (iii) it has 24$\mu$m emission as seen in Fig.\,\ref{fig:ysothumb}. The source has marked PolyAromaticHydrocarbon (PAH) emission from the 8.6$\mu$m feature. 

\begin{figure}
	\includegraphics[width=\columnwidth]{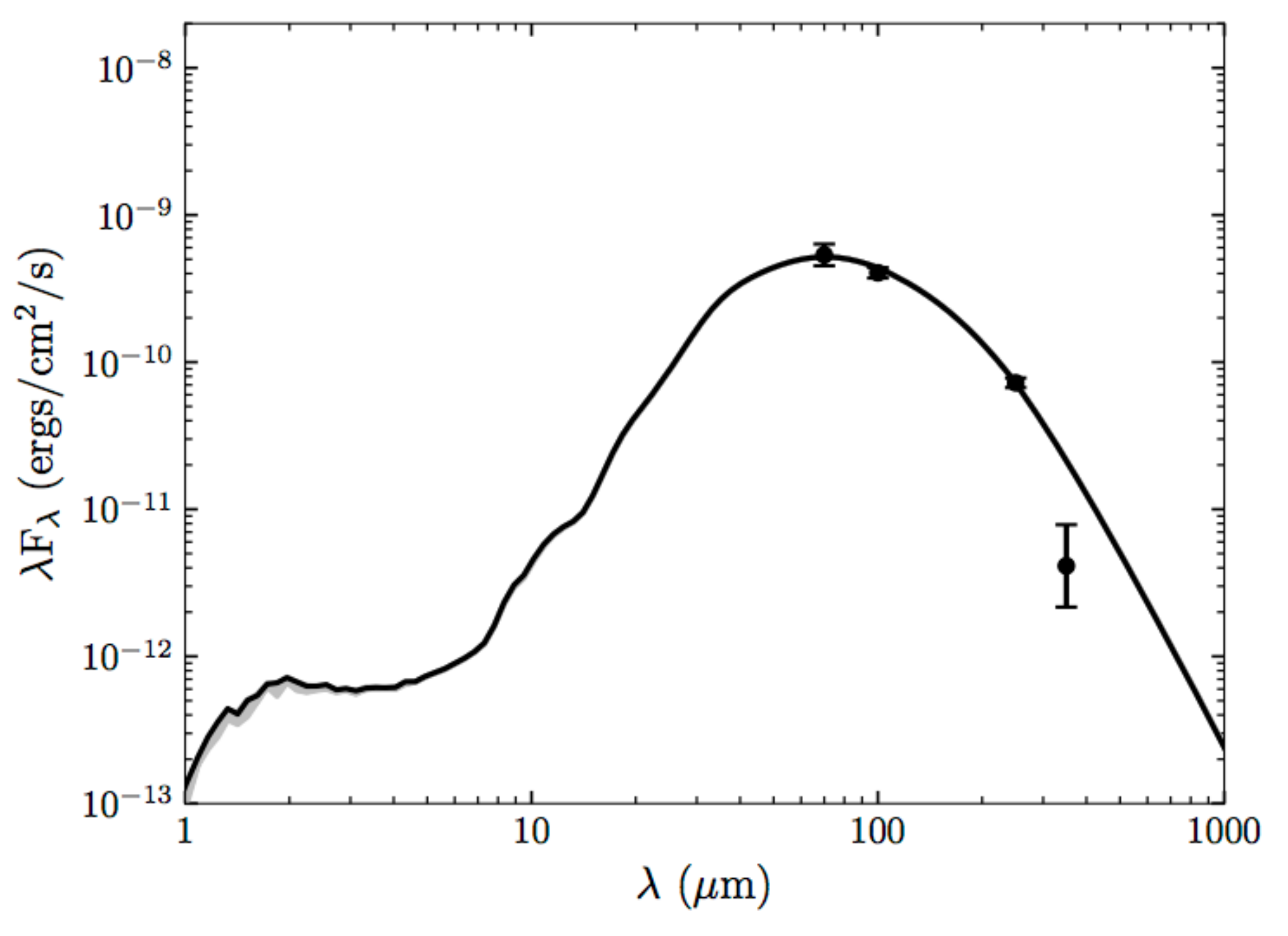}
    \caption{SED of KN2-B with best-fit models from Robitalle (2017), overlaid with {\it Herschel} 100-350$\mu$m photometry and corresponding error bars. The best fit model (with $\chi^2$/datapoint$<$3) with a stellar temperature of $\sim$18500\,K is shown as a solid black line. Other models with $\chi^2$/datapoint$<$5 are also shown as solid grey lines.}
    \label{fig:sed}
\end{figure}

We achieve an excellent fit for KN2-B with the SED models of Robitalle et al. (2017). The best fit has a $\chi^2$/datapoint$<$3 and is shown in Fig.\,\ref{fig:sed}. The parameters of the best fit model were a stellar temperature of 18500$\pm$1000 K, which translates to a stellar mass exceeding 20\,$M_{\odot}$, with the $\dot M_{\rm env}/M_{\ast}$ from the model fit suggestive of a Class 0 massive YSO (Fig.\,\ref{fig:sed}).The temperature estimated from the blackbody fit is 28.8\,$\pm$4\,K, with the integrated logarithm of luminosity log\,$L/L_{\odot}$=4.59 which further indicates a Class 0 classification (Andr{\'e} et al. 2010). The luminosity of the source is much higher than expected for starless clouds heated by the ambient radiation field, considering which it is likely a true high-mass YSO (Seale et al. 2014). The estimated extinction from the SED fit (of $A_V$=6.9) suggests that the candidate protostar is shielded from external photoionization. Moreover, the external heating at wavelengths longwards of 100$\mu$m does not affect significantly the SED (Pavlyuchenkov et al. 2012). Finally, as noted in Sec.\,5.1, the parameters estimated from the SED fits with the Robitalle (2017) models may be ambiguous due to the limited range of parameters considered, but they do provide a crude guide to source classification. We also stress that KN2-B is classified as a high-mass YSOs based on the independent schemes of Seale et al. (2014) based on source intensity and morphology; and the SED models of Robitalle et al. (2017) based on the flux distribution of the source in the fIR.  The age of KN2-B from the model fit is $\lesssim$\,0.1\,Myr, and is smaller by an order of magnitude than the crossing timescale of the cloud (the crossing timescale is the length over the velocity) of 0.66\,Myr. 

KN2-A is likely a Class II source based on its {\it Spitzer} colours. The $\alpha_{\rm SED}$ slope derived from the 3.6-8$\mu$m photometry is 1.22$\pm$0.2, which falls into the Class II category (Greene et al. 1994). Our classification as a Class II YSO from the SED slope agrees with the Gruendl \& Chu (2009) classification based on the position of the KN2-A in {\it Spitzer} colour-colour and colour-magnitude diagrams. We note the there is no 24$\mu$m counterpart to KN2-A, and it also does not have a {\it Herschel} counterpart according to the procedure followed Seale et al. (2014), who cross match {\it Spitzer} and {\it Herschel} catalogues in the LMC. They adopt a cross match radius of 0.5$\times$FWHM to cross match the two catalogues in crowded regions. If we still consider that some of the fIR flux from KN2-A is assigned towards KN2-B, we find that $\gtrsim$5\% of the fIR flux of KN2-B is required to change the classification of KN2-A from Class II to Class I. There is no high mass Class II YSOs population on the northern side. 

\begin{table*}
	\centering
	\caption{Candidate Class 0/I YSOs from {\it Spitzer} and {\it Herschel} photometry.}
	\label{tab:example_table}
\begin{tabular}{lcccccc}        
\hline\hline                 
ID & Wavelength\,($\mu$m) & R.A. (J2000) & Dec. (J2000) & Class & $\alpha_{\rm SED}$ & Comments\\    
\hline 
KN2-A & 1.2--8 & 05$^h$38$^m$39$^s$7 & $-$69$^{\rm \circ}$05$'$38.1$''$ & II & 1.22$\pm$0.2 & S5 in Walborn et al. (2013) \\
KN2-B & 100-350 & 05$^h$38$^m$39$^s$1 & $-$69$^{\rm \circ}$05$'$33.8$''$ & 0 & -- & $T_{\rm b}$=29$\pm$4\,K; point like in 24$\mu$m imaging  \\
KN3-A & 3.6-500 & 05$^h$38$^m$36$^s$3 & $-$69$^{\rm \circ}$05$'$24$''$ & Dust(?) & 2.65$\pm$0.7 & $T_{\rm b}$=31.5$\pm$7\,K; extended in 24$\mu$m imaging  \\
 &  &  &  &  &  & 0.9$''$ between {\it Herschel} 100$\mu$m\& {\it Spitzer} source  \\
\hline                        
\end{tabular}\\
\end{table*}

\subsubsection{KN3}

KN3-A is a bright {\it Herschel} source and is extended in 100--350$\mu$m imaging. The {\it Herschel} detection lies within 0.9$''$ of a Gruendl \& Chu (2009) {\it Spitzer} source. We consider the mIR and fIR detection to be of the same source, as they lie well within the FWHM of the {\it Herschel} and {\it Spitzer} imaging. Gruendl \& Chu (2009) classified the source as likely originating from dust, and we see that their exists no point-like or otherwise 24$\mu$m counterpart. The whole SED fits rather poorly with the models of Robitalle et al.  (2017), with the best fits achieved when discarding the 4.5$\mu$m photometry of $\chi^2$/datapoint $>$100. Examination of the 24$\mu$m imaging reveals no distinct point-like source (Fig.\,\ref{fig:ysothumb}; right dashed circle). Given the flux at long wavelengths, and the visible extended emission in the imaging we cannot classify the source as a YSO conclusively. A modified blackbody fit suggests a peak temperature of 31.5$\pm$6\,K, with the integrated total luminosity slightly less than 10$^4$\,$L_{\odot}$. This lies at the boundary of the classifying scheme between interstellar starless clouds heated by the ambient radiation field (Seale et al. 2014), and YSOs. The ambiguous classification, and lack of any clear point-like source in the nIR and fIR images suggests that this source is likely a starless dust cloud, probably heated by the ambient ionizing field. 

A comparison of the multi-wavelength images in Fig.\,\ref{fig:ysothumb} demonstrates the differences between KN2-B and KN3-A. At 24$\mu$m, diffuse emission in star-forming regions is likely to originate from heated gas from inner regions of protostars (Chambers et al. 2009). In KN2-B, the 24$\mu$m emission is clearly detected at 3$\sigma$ level with respect to the immediate surroundings. In KN3-A, 24$\mu$m emission above the 3$\sigma$ threshold is not detected. At 100$\mu$m, and 160$\mu$m, both sources appear as point sources. But, at longer wavelengths, the KN3-A source is extended or disappears (see the 250$\mu$m image), suggesting that it is more likely to be dust, rather than a protostellar candidate when compared to the nature of the source detected at the position of KN2-B in the 250$\mu$m image.

\section{A picture of the stapler nebula in R136}

\begin{figure*}
	\plotone{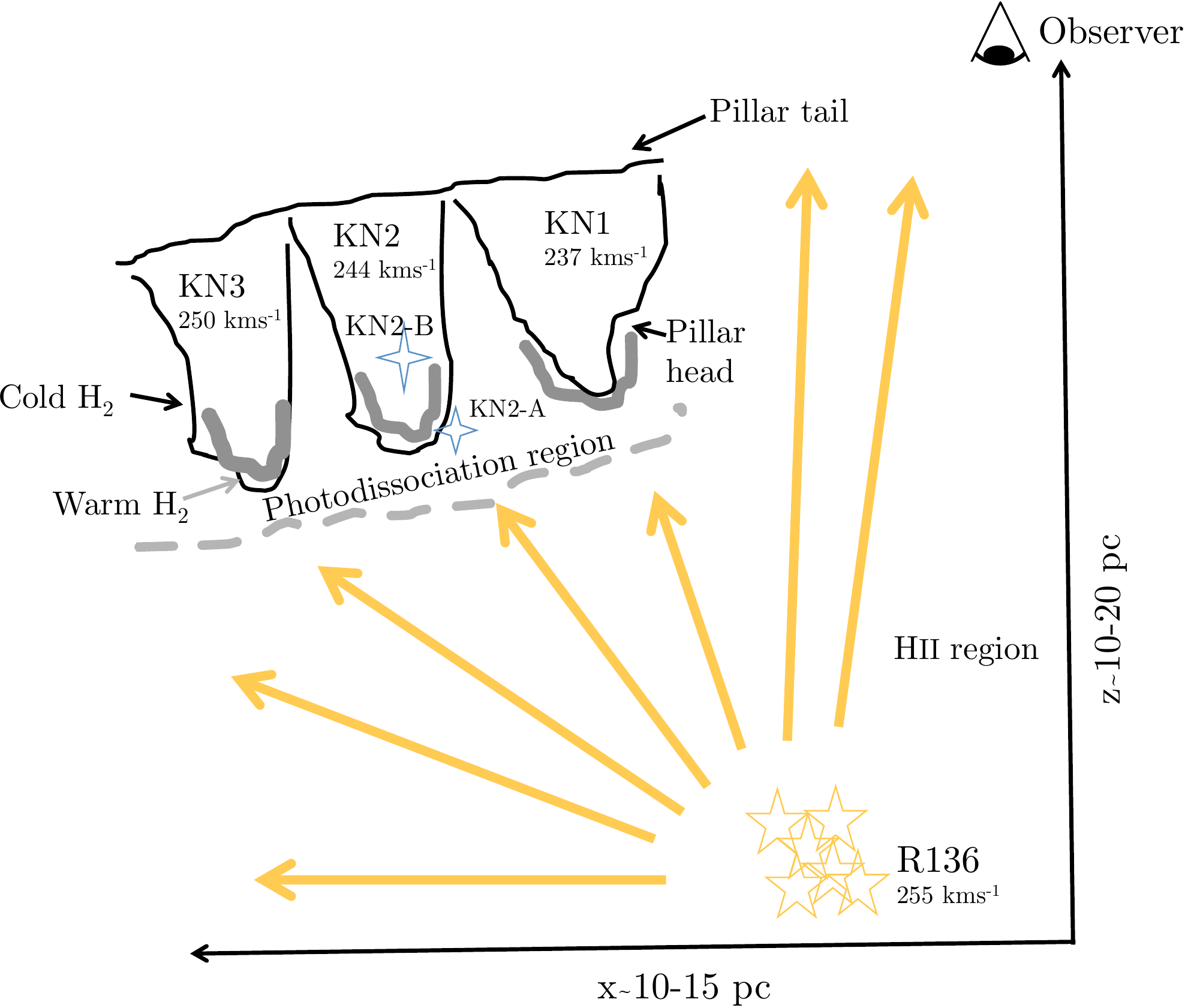}
    \caption{Schematic of the stapler nebula. The model is flattened in the y direction, with the direction of the observer in z-axis indicated. Ionising radiation from the massive R\,136 stars (yellow asterisks, with the mean radial velocity given in the local standard of rest frame) photoionises (given by yellow arrows) the blue shifted cold molecular gas lying in pillar-like features (solid lines), leading to the observed CO\,(2-1) line widths. This leads to a photoionisation region in the head of the pillar from where the Br$\gamma$ emission arises (grey thick lines). The tail of the pillar is comparatively colder, where the cold molecular gas is detected. The individual molecular clouds are identified along with their central velocities in the local standard of rest frame. In KN2, we detect the candidate protostar KN2-B (given by the large blue star), which is likely a forming high mass star. The previously known YSO candidate KN2-A is also shown as a small blue star.}
    \label{fig:model}
\end{figure*}

From our deep CO\,(2-1) observations, we identified three molecular clouds separated in velocity we name KN1-3 lying within 14\,pc in projection; and $\lesssim$20\,pc in the line of sight from the YMC R\,136. These clouds lie north of R\,136 and stretch for 14\,pc in projection from southeast to northwest. This axis is perpendicular to the previously identified giant molecular clouds by Johansson et al. (1998). The $V_{\rm lsr}$ of the clouds ranges from 237\,km\,s$^{-1}$ for the cloud directly above R\,136 in projection to 250\,km\,s$^{-1}$ for the furthest cloud, which is similar to the mean radial velocity of the stellar population in R\,136. These clouds display relatively large linewidths for their radii, lying above the predicted relation for clouds in virial equilibrium according to Larson's first law. By plotting ${\sigma_{v}}^2$/$r$ of the KN clouds as a function of their $\Sigma_{\rm H_{2}}$, we show that KN clouds depart significantly from the relation for virialized, isolated clouds confined by self-gravity. The KN clouds are most likely confined by external pressures upwards of $\sim$\,10$^6$\,cm$^{-3}$\,K in this scenario. Nonetheless, the results from our CO\,(2-1) line observations show that there are three molecular clouds lying near R\,136, whose $V_{\rm lsr}$ increase as a function of projected distance from the cluster. The resolved clouds are not in virial equilibrium, but are highly turbulent requiring external pressure to confine them.

Br$\gamma$ and H$_2$ nIR emission line imaging reveal that the CO clouds coincide with the dense H$_2$ clumpy structures, while Br$\gamma$ is diffuse and exhibits little spatial coincidence with the CO clouds. The lack of background stars at optical wavelengths (Fig.\,1) suggests that these clouds lie in front of the cluster. Based on the H$_2$/Br$\gamma$ ratio, we suggest that these clouds are UV-heated by stellar radiation emanating from R\,136. These results suggests that both the H$_2$ nIR emission, and CO emission arise from molecular clouds near R\,136. The molecular clouds lie in front of R\,136 from our perspective. The backside of the molecular clouds is being ionised by R\,136 leading to diffuse Br$\gamma$ emission and weak H$\alpha$ emission. The excited H$_2$ emission coincides with the peak of CO emission further giving weight to our view of photodissociated molecular clouds near R\,136. 

Finally, a search for results of on-going star formation yields interesting results. KN1 shows no signs of active star formation. In fIR {\it Herschel} 100-350$\mu$m photometry, we identify a likely Class 0/I object KN2-A corresponding to the peak of the molecular cloud KN2. It is known from previous CS observations that the densities near this source exceed $n>$10$^6$. Near this source (within 5$''$) lies a Class II object KN2-A visible in {\it Spitzer} imaging, which has been classified as a YSO by Gruendl \& Chu (2009) and Walborn et al. (2013). Towards KN3, we detect a {\it Herschel} 100-160$\mu$m source (KN3-A) close to a {\it Spitzer} source identified by Gruendl \& Chu (2009). However, inspection of both the {\it Spitzer} and {\it Herschel} imaging suggests KN3-A is likely dusty in nature, which agrees with the Gruendl \& Chu (2009) classification. 

The complete scenario can be visualised in the toy model shown in Fig.\,\ref{fig:model} along the observer's viewpoint. The natal molecular cloud of R\,136 has an arc like structure, protruding in front of it along our line of sight, and lying directly above it and stretching towards the north west in projection forming an arc shaped structure. The structure may have been carved out from an initially spherical cloud. After the formation of the cluster, the natal molecular cloud has been steadily ionised giving rise to excited H$_2$ emission. The excited boundary lies on the backside of the molecular cloud from the observer's viewpoint. The radiation from R\,136 is eroding the molecular clouds, possibly leading to the formation of structures similar to the ``pillars of creation" observed in galactic regions such as M\,16. The observer may be visualising the tail of such pillars face on. The observed high velocity line widths could then be explained as a manifestation of the expected velocity differences between the head and tail of a photoevaporating cloud. Alternatively, the results also support the picture that the KN clouds may likely be confined by external pressures.  High angular resolution IFU spectroscopy of the clouds which will reveal their gas kinematics and ionization structure (McLeod et al. 2015) can help differentiate between the two proposed scenarios. Massive star formation is occurring in the middle molecular cloud or pillar, while the remaining two molecular clouds show no active signs of massive star formation.

\section{Summary and future work}

Our results present a picture of three CO molecular clouds separated in velocity lying $\lesssim$20\,pc in front of R\,136, and between 2--14\,pc away in projection from the cluster. We appear to view the tail of pillar-like structures whose ionised heads are pointing towards R\,136. The observed high line widths of the molecular clouds for their sizes with respect to the canonical $\sigma_{v}$-$r$ relation are likely due to the gas being pushed away from the head of the pillar-like structures by the ionization radiation produced by the massive stars in R\,136, or due to external pressure confining the clouds. A massive YSO (KN2-B) is detected inside the KN2 molecular cloud, indicating active star formation. These results suggest that 1.5--3\,Myr R\,136 cluster is in the process eroding the KN molecular clouds. During this process a new generation of stars is able to form from the reservoir of cold molecular gas that is in the process of being destroyed by stellar feedback in structures that may be similar to the pillars of creation seen in Galactic regions such as M\,16 and NGC\,3603 from a different perspective.

We have demonstrated that with sensitive CO\,(2-1) observations, molecular clouds of $\sim$10$^4$\,$M_{\odot}$ exist surprisingly close to the YMC R\,136. This observation challenges the results of most simulations of feedback from massive stars in YMC (see Dale 2015 for a review), where feedback is able to successfully expel gas close to YMCs within a few Myrs. Future high angular resolution sensitive observations of multiple molecular lines using the current generation of interferometers (for e.g. using the Atacama Large Millimeter/submillimeter Array (ALMA) which can achieve a resolution comparable to optical/nIR imaging in CO lines) may reveal in exquisite detail the structure, and densities of these clouds. Combined with observations of the photodissociation region from fine structure lines (Chevance et al. 2016), one can gather enough information about the densities, pressures, and radiation field acting on the KN clouds and the H{\scriptsize II} region to predict if the clouds are dense and massive enough to survive photoionisation and form a new massive stellar cluster, or whether star formation will be abruptly terminated due to feedback. In a broader sense, sensitive radio observations in CO lines may reveal previously undetected molecular clouds lying close to YMCs, that may nurture a new generation of star formation.

\acknowledgments
The authors thank Jes{\'us} Ma{\'i}z Apell{\'a}niz for kindly providing us with the {\it HST} mosaic, and sharing with us a pictorial etymology of the stapler nebula. We thank Sherry Yeh for generously providing her Br$\gamma$ image, and Hugo Saldano for his help in visualizing the radio data. The anonymous referee is thanked for providing constructive comments, and in helping to improve the overall impact of the paper. V.M.K. acknowledges support from the FONDECYT-CHILE Fellowship grant N$^{\rm o}$3116017. M.R. acknowledges support from CONICYT (CHILE) through FONDECYT grant N$^{\rm o}$1140839, and partial support through BASAL\,PFB-06. This work made use of the CLASS software, and the Python packages ApLpy, Numpy, Scipy and Matplotlib for analysis and presentation.
%

\vspace{5mm}

\end{document}